\documentclass{osa-article1}

\journal{oe}
\newcommand\beq{\begin{equation}}
\newcommand\eeq{\end{equation}}

\usepackage{graphicx}% Include figure files
\usepackage{dcolumn}% Align table columns on decimal point
\usepackage{bm}% bold math

\begin{document}

\title{Effects of deterministic disorder at deeply subwavelength scales in multilayered dielectric metamaterials}

\author{Marino Coppolaro, Giuseppe Castaldi, and Vincenzo Galdi\authormark{*}}

\address{Fields \& Waves Lab, Department of Engineering, University of Sannio, I-82100, Benevento, Italy}

\email{\authormark{*}vgaldi@unisannio.it} %% email address is required



%%%%%%%%%%%%%%%%%%%% Created:			  11/11/2019
%%%%%%%%%%%%%%%%%%%% Last revised:		03/03/2020

\begin{abstract}
It is common understanding that multilayered dielectric metamaterials, in the regime of {\em deeply subwavelength} layers, are accurately described by simple effective-medium models based on mixing formulas that {\em do not} depend on the spatial arrangement. In the wake of recent studies that have shown counterintuitive examples of periodic and aperiodic (orderly or random) scenarios in which this premise breaks down, we study here the effects of {\em deterministic disorder}. With specific reference to a model based on Golay-Rudin-Shapiro sequences, we illustrate certain peculiar boundary effects that can occur in finite-size dielectric multilayers, leading to anomalous light-transport properties that are in stark contrast with the predictions from conventional effective-medium theory. Via parametric and comparative studies, we elucidate the underlying physical mechanisms, also highlighting similarities and differences with respect to previously studied geometries. Our outcomes may inspire potential applications to optical sensing, switching and lasing.
\end{abstract}

%%%%%%%%%%%%%%%%%%%%%%%%%%%%%%%%%%%%%%%%%%%%%%%%%%%%%%%%%%%%%%
\section{Introduction}
%%%%%%%%%%%%%%%%%%%%%%%%%%%%%%%%%%%%%%%%%%%%%%%%%%%%%%%%%%%%%%
Metamaterials \cite{Capolino:2009vr} are typically described in terms of {\em effective} constitutive parameters (e.g., permittivity and permeability) that generally depend on the material and geometric properties of their constituents, as well as their spatial arrangement.  This {\em homogenization} process is crucial for the characterization, modeling and phenomenological understanding, and can be carried out in various ways, possibly taking into account the spatial-dispersion effects (see, e.g., \cite{Silveirinha:2007mh,Alu:2011fp,Chebykin:2011ne,Chern:2013sd,Ciattoni:2015nh} and references therein for a sparse sampling). In the presence of {\em deeply subwavelength} dielectric inclusions, 
spatial-dispersion effects are usually very weak, and
the effective properties primarily depend on the
shapes, orientations, and filling fractions of the constituents, whereas their actual dimensions and spatial arrangement play a negligible role. In these conditions,
{\em local} effective-medium-theory (EMT) approaches relying on mixing formulas (e.g., Maxwell-Garnett, Bruggeman) are typically sufficient to accurately capture the optical response \cite{Sihvola:1999em}. 

In connection with multilayered dielectric metamaterials, a series of recent studies initiated by Herzig Sheinfux {\em et al.} \cite{Sheinfux:2014sm} have demonstrated the existence of critical parameter regimes where the simple EMT approximation breaks down, in spite of the deeply subwavelength size of the layers. This phenomenon, manifested as significant differences between the actual and EMT-predicted optical responses of {\em finite-thickness} samples for specific incidence directions, is attributable to {\em boundary effects} induced by an unusual transport mechanism which mixes evanescent and propagating characteristics \cite{Sheinfux:2014sm}. In addition to the experimental evidence \cite{Zhukovsky:2015ed}, subsequent studies \cite{Andryieuski:2015ae,Popov:2016oa,Lei:2017rt,Maurel:2018so,Castaldi:2018be,Gorlach:2019bc} have provided further insights and put forward alternative models and possible (e.g., nonlocal, magneto-electric) extensions to capture these counterintuitive effects. 

Interestingly, these phenomena are not limited to periodic arrangements. In particular, anomalous Anderson localization was observed theoretically \cite{Sheinfux:2016cr} and experimentally \cite{Sheinfux:2017oo} in  {\em randomly disordered} dielectric multilayers, once again in a deeply subwavelength regime where conventional EMT predicts instead a moderately transmissive response. More recently, inspired by the permeating concept of ``quasicrystals'' \cite{Macia:2006tr,DalNegro:2011da},
we have started exploring {\em aperiodically ordered} multilayer geometries, lying in between perfect periodicity and random disorder. With specific reference to the Thue-Morse geometry \cite{Coppolaro:2018ao}, we have shown the emergence of the EMT-breakdown phenomenon at deeply subwavelength scales, but with mechanisms and footprints (e.g., fractal gaps, quasi-localized states)  that are distinctive of aperiodic order and
differ fundamentally from those observed in the periodic and random scenarios.

It is worth remarking that, far from being a physical oddity of purely academic interest, the enhanced sensitivity to changes of features on a deeply subwavelength scale may find intriguing applications to optical sensing, switching and lasing. It is therefore interesting to explore different scenarios and geometries in order to gain further insight in the underlying mechanisms, as well as to unveil new effects and critical parameter ranges. While it is not possible to capture in a single example the wealth of  complex features and phenomena that characterize the realm of ``orderly disorder'', there are certain prototypical geometries that are especially interesting and representative.

Within this framework, we study here multilayered dielectric metamaterials based on the Golay-Rudin-Shapiro (GRS) sequences \cite{Golay:1951sm,Shapiro:1952ep,Rudin:1959st}, which constitute one of the simplest examples of  {\em deterministic disorder}. Specifically, after a brief outline of the problem  (Sec. \ref{Sec:Statement}), we present a series of parametric studies (Sec. \ref{Sec:Results}) which indicate the emergence of peculiar boundary effects leading to anomalous light-transport properties (localization, field enhancement, absorption, and lasing) that are not predicted by conventional EMT and are also not observable in periodic counterparts. 
Finally, we provide some brief concluding remarks and hints for possible applications and future research (Sec. \ref{Sec:Conclusions}).

%%%%%%%%%%%%%%%%%%%%%%%%%%%%%%%%%%%%%%%%%%%%%%%%%%%%%%%%%%%%%%
\section{Problem outline}
%%%%%%%%%%%%%%%%%%%%%%%%%%%%%%%%%%%%%%%%%%%%%%%%%%%%%%%%%%%%%%
\label{Sec:Statement}

%############################################################
%                Figure 1
%
\begin{figure}
	\centering\includegraphics[width=12cm]{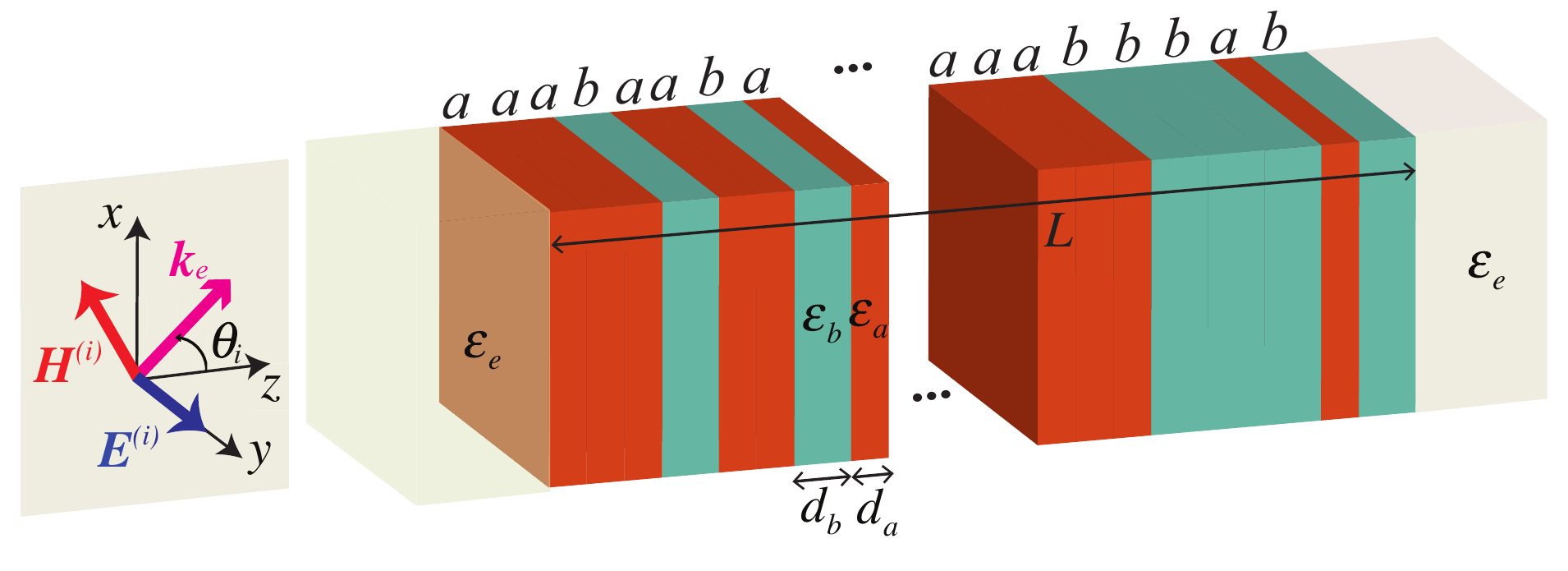}
	\caption{Problem schematic (details in the text).}
	\label{Figure1}
\end{figure}
%############################################################

%===============================================================================================
\subsection{Geometry}
%===============================================================================================
Referring to Fig. \ref{Figure1} for illustration, 
we consider a multilayered  metamaterial featuring two types (``$a$'' and ``$b$'') of dielectric constituent layers 
with relative permittivities $\varepsilon_a$ and $\varepsilon_b$, and thickness $d_a$ and $d_b$, respectively, embedded in a homogeneous dielectric background with relative permittivity $\varepsilon_e$. All materials are assumed as nonmagnetic (relative permeability $\mu=1$).
The layers are assumed of infinite extent in the $x-y$ plane, and are stacked aperiodically along the $z$-direction following a deterministic rule based on the GRS sequences \cite{Golay:1951sm,Shapiro:1952ep,Rudin:1959st}, i.e.,  two-symbol aperiodic sequences characterized by deterministic disorder.
Among the various generation algorithms available, we consider the so-called  GRS polynomials \cite{Golay:1951sm,Shapiro:1952ep,Rudin:1959st}, defined recursively by the intertwined relationships
\begin{subequations}
\begin{eqnarray}
P_{\nu+1}\left(X\right)&=&P_{\nu}\left(X\right)+X^{2^\nu}Q_{\nu}\left(X\right),\\
Q_{\nu+1}\left(X\right)&=&Q_{\nu}\left(X\right)-X^{2^\nu}P_{\nu}\left(X\right),
\end{eqnarray}
\end{subequations}
initialized with $P_0=Q_0=1$. It can be verified that $P_{\nu}$ and $Q_{\nu}$ have $N=2^\nu$ coefficients ($p_n$ and $q_n$, respectively) with value $\pm 1$, which can also been calculated explicitly as \cite{Rudin:1959st}
\begin{subequations}
\beq
p_0=1,\quad p_{2n}=p_n,\quad p_{2n+1}=\left(-1\right)^np_n,
\eeq
\beq
q_n=\left\{
\begin{array}{ll}
	p_n,\quad n=0,...,\frac{N}{2}-1,\\
	-p_n,\quad n=\frac{N}{2},...,N-1.
	\end{array}
\right.
\eeq
\end{subequations}
Our multilayer geometry is obtained from these coefficients, by mapping the binary alphabet $\left\{-1,1\right\}$ onto $\left\{a,b\right\}$, and stacking the layers accordingly. 
In spite of their fully deterministic character, GRS sequences exhibit statistical and spectral properties that are more akin to random sequences \cite{Fogg:2002si,Wolff:2017ds}. First, it can be 
shown that, in the asymptotic limit $N\rightarrow\infty$, the statistical frequencies of occurrence of the two symbols, $f_a$ and $f_b$, are identical (i.e., $f_a=f_b=0.5$) \cite{Berthe:1994cp}. Specifically, for $Q_{\nu}$ polynomials with odd $\nu$, this property is valid irrespective of the sequence length. It can also be shown that the maximum number of consecutive symbols is four \cite{Wolff:2017ds}. 
More interestingly, the correlation properties of GRS sequences closely resemble those of white-noise-like random sequences \cite{Fogg:2002si}. Their Fourier spectra are devoid of sharp Bragg-type peaks and exhibit an {\em absolutely continuous} character \cite{Fogg:2002si} with extremal properties closely related with the quest for ``flat polynomials'' in harmonic analysis \cite{Rudin:1959st}. Accordingly, alternative terms such as ``flat-spectrum aperiodic order'' or ``pseudo-randomness'' are also utilized in the topical literature to define this type of geometries.

So far, in electromagnetics and optical engineering, GRS sequences have been explored in connection with antenna arrays \cite{Galdi:2005rp}, spread-spectrum communications and encryption \cite{laCour-Harbo:2008ot},  diffuse scattering \cite{Moccia:2017cm,Moccia:2018sc},
nanoplasmonic arrays \cite{Gopinath:2009da,Forestiere:2009tr,Boriskina:2010fc}, and photonic crystals  \cite{Axel:1992oc,Vasconcelos:1999tf,Hiltunen:2007mo,Agarwal:2009op,Bouazzi:2012of,Trabelsi:2016ns}.

%===============================================================================================
\subsection{Statement}
%===============================================================================================
As in previous studies on periodic \cite{Castaldi:2018be} and aperiodically ordered \cite{Coppolaro:2018ao} dielectric multilayers, we assume a plane-wave illumination, with transverse-electric (TE) polarization, incidence angle $\theta_i$ and suppressed $\exp\left(-i\omega t\right)$ time-harmonic  dependence, characterized by a unit-amplitude $y$-directed electric field (see Fig. \ref{Figure1})
\beq
E_y^{(i)}\left(x,z\right)=\exp\left[i\left(k_x x+k_{ze} z\right)\right],
\eeq
and wavevector ${\bf k}_e\equiv\left(k_x,k_{ze}\right)=\left(k \sqrt{\varepsilon_e}\sin\theta_i,k \sqrt{\varepsilon_e}\cos\theta_i\right)$, with $k=\omega/c=2\pi/\lambda$ denoting the vacuum wavenumber (and $c$ and $\lambda$ the corresponding wave velocity and wavelength, respectively).

Previous studies on GRS-type multilayered dielectric structures \cite{Axel:1992oc,Vasconcelos:1999tf,Hiltunen:2007mo,Agarwal:2009op,Bouazzi:2012of,Trabelsi:2016ns} have focused on the {\em diffractive} (photonic-crystal) regime characterized by {\em moderately thick} layers, showing interesting light-transport properties in terms of bandgap, filtering and field-enhancement.  Here, we focus instead on the insofar unexplored  {\em deeply subwavelength} regime, 
characterized by $d_{a,b}\ll \lambda$, in which the optical properties of such metamaterials are typically well described by simple EMT models based on Maxwell-Garnett-type mixing formulas \cite{Sihvola:1999em}. For the assumed TE polarization, such effective medium is characterized by only one relevant relative-permittivity component \cite{Sihvola:1999em},
\beq
{\bar \varepsilon}_{\parallel}=\frac{f_a\varepsilon_a d_a+f_b\varepsilon_b d_b}{f_ad_a+f_bd_b}=\frac{\varepsilon_a d_a+\varepsilon_b d_b}{d_a+d_b},
\label{eq:EMT}
\eeq
where we have exploited the aforementioned property $f_a=f_b$ of GRS-type sequences. 
It is worth stressing that the expression in (\ref{eq:EMT}) depends {\em only} on the material properties and proportions of the two constituents, and {\em not} on the actual layer thicknesses and their spatial arrangement. In other words, at deeply subwavelength scales, different geometrical arrangements (periodic or aperiodic, deterministic or random) of the layers should not affect the optical response, as long as the overall proportions of the constituents is maintained. Contrary to this conventional supposition, in what follows, we show that there exist certain critical parameter regimes where the GRS-type deterministic disorder, even though at deeply subwavelength scales, induces anomalous light-transport properties that are not predicted by the EMT model and are also markedly different from those observed in the periodic case \cite{Sheinfux:2014sm,Andryieuski:2015ae,Popov:2016oa,Lei:2017rt,Maurel:2018so,Castaldi:2018be,Gorlach:2019bc}. In order to ensure that the expression of the effective parameter in (\ref{eq:EMT}) is always exactly valid, we restrict our attention to GRS sequences generated via $Q_{\nu}$-type polynomials with odd $\nu$, for which $f_a=f_b$ any stage of growth (not only asymptotically).

%############################################################
%                Figure 2
%
\begin{figure}
	\centering\includegraphics[width=13.2cm]{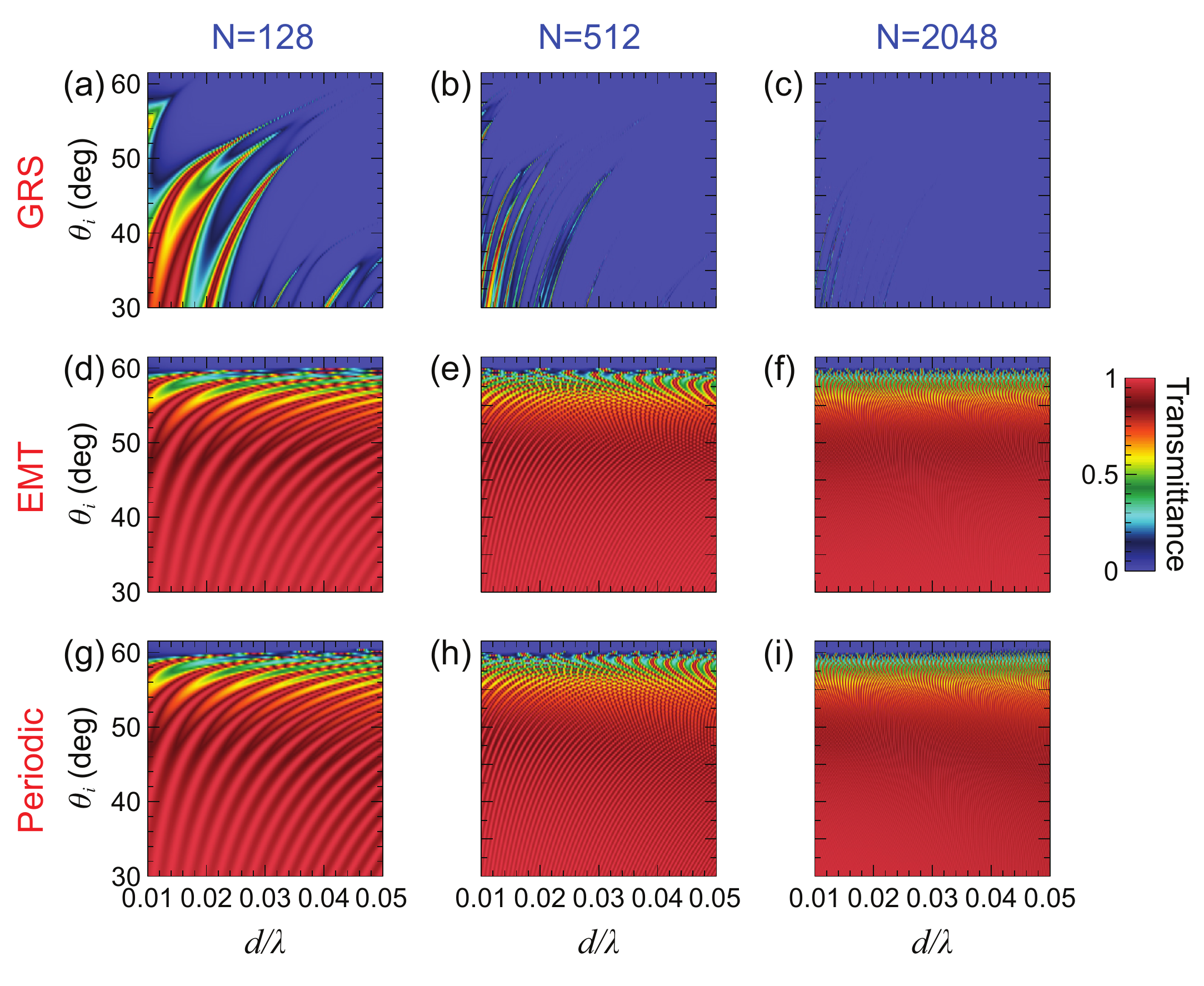}
	\caption{Comparison between the transmittance responses of multilayered dielectric metamaterials with different geometrical arrangements, for $\varepsilon_a=1$, $\varepsilon_b=5$, $d_a=d_b=d$, $\varepsilon_e=4$, as a function of the layer electrical thickness $d/\lambda$ and incidence angle $\theta_i$.
		(a)--(c) $Q_{\nu}$-type GRS geometries at stages of growth $\nu=7$ ($N=128$ layers), $\nu=9$ ($N=512$ layers), and $\nu=11$ ($N=2048$ layers), respectively.  (d)--(f) Corresponding EMT predictions (${\bar \varepsilon}_{\parallel}=3$). (g)--(i) Corresponding responses for periodic counterparts.}
\label{Figure2}	
\end{figure}
%############################################################

%%%%%%%%%%%%%%%%%%%%%%%%%%%%%%%%%%%%%%%%%%%%%%%%%%%%%%%%%%%%%%
\section{Results and discussion}
%%%%%%%%%%%%%%%%%%%%%%%%%%%%%%%%%%%%%%%%%%%%%%%%%%%%%%%%%%%%%%
\label{Sec:Results}

%===============================================================================================
\subsection{Generalities}
%===============================================================================================
As in previous studies on periodic and aperiodic (random or ordered) configurations \cite{Sheinfux:2014sm,Sheinfux:2016cr,Sheinfux:2017oo,Castaldi:2018be,Coppolaro:2018ao},
we assume for the exterior background medium $\varepsilon_e=4$, and for the material layers $\varepsilon_a=1$ and $\varepsilon_b=5$, with same layer thickness ($d_a=d_b=d$), neglecting for now the presence of losses. From (\ref{eq:EMT}), this yields ${\bar \varepsilon}_{\parallel}=3$ for the effective medium. Moreover, we focus on the deeply subwavelength parameter regime $0.01\lambda<d<0.05\lambda$, and on incidence directions $30^o<\theta_i\lesssim 60^o$. For the assumed parameters, this angular-incidence range guarantees that the field is evanescent in the ``$a$''-type layers and propagating in the ``$b$''-type ones. The regime characterized by smaller angles of incidence ($0<\theta_i< 30^o$), in  which the field is propagating in both layers, has been extensively investigated in previous studies on photonic crystals \cite{Axel:1992oc,Vasconcelos:1999tf,Hiltunen:2007mo,Agarwal:2009op,Bouazzi:2012of,Trabelsi:2016ns}, and is not of interest here. Overall, the assumed parameter regime is rather unusual, as the phase-accumulation mechanism is essentially dominated by the discrete jumps (Fresnel-type reflection/transmission) at the layer interfaces rather than the delay acquired via propagation through the (deeply subwavelength) layers. In previous studies on periodic and aperiodic structures \cite{Sheinfux:2014sm,Sheinfux:2016cr,Sheinfux:2017oo,Castaldi:2018be,Coppolaro:2018ao}, it was shown that such mechanism can lead to the buildup of boundary effects that can substantially enhance the (otherwise negligible) nonlocal effects, thereby yielding strong departures of the optical response from the EMT predictions. These effects tend to be particularly enhanced nearby the critical angle defining the total-internal-reflection condition for the effective medium, ${\bar \theta}_c=\arcsin\left(\sqrt{{\bar \varepsilon}_{\parallel}/\varepsilon_e}\right)=60^o$.

To illustrate the onset of the aforementioned boundary effects, in what follows, we compare systematically the optical responses of GRS-type multilayered metamaterials (at various stages of growth) with the corresponding EMT predictions. Moreover, to gain some insight in the underlying physical mechanisms, we also compare these results with periodically arranged benchmarks with same effective properties. Our numerical simulations below are carried out via straightforward implementation of the transfer-matrix approach \cite{Born:1999un}, which provides a rigorous solution of the problem.

%############################################################
%                Figure 3
%
\begin{figure}
	\centering\includegraphics[width=13.2cm]{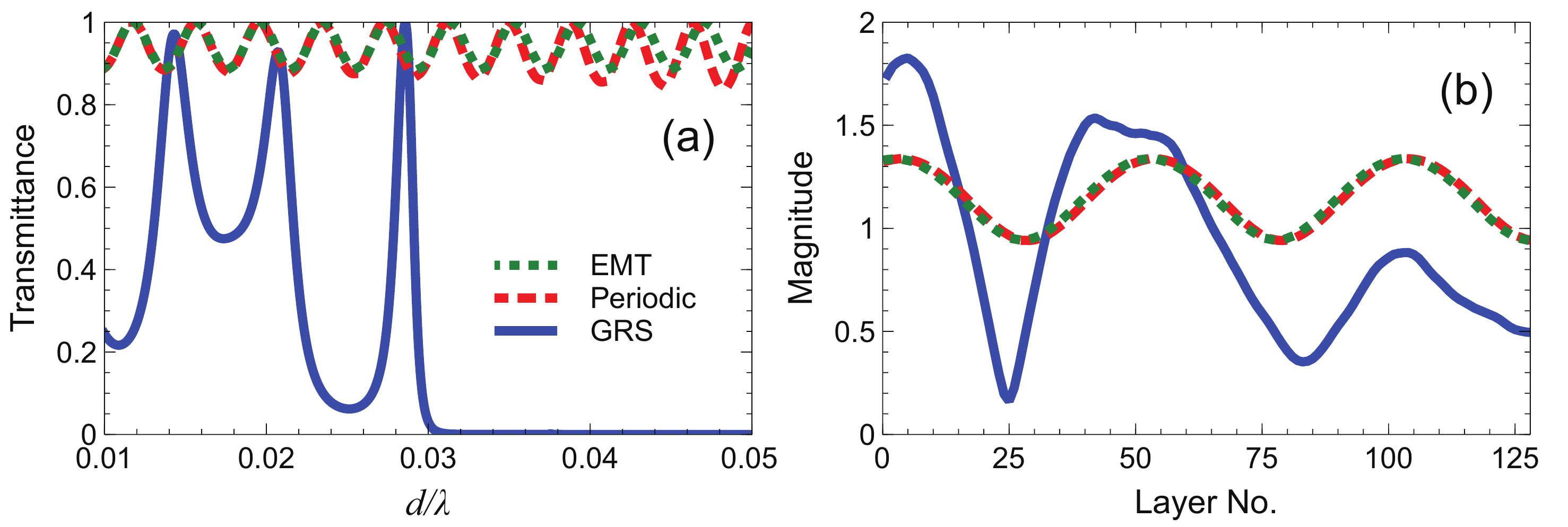}
	\caption{(a) Representative transmittance cuts from Figs. \ref{Figure2}a (GRS; blue-solid), \ref{Figure2}d (EMT; green-dotted), and \ref{Figure2}g (periodic; red-dashed), for $\theta_i=45.1^o$. (b) Corresponding electric-field (normalized-magnitude) distributions for $d/\lambda=0.01$.}
	\label{Figure3}	
\end{figure}
%############################################################

%===============================================================================================
\subsection{Transmission and localization}
%===============================================================================================
Figure \ref{Figure2} shows a comparison among the transmittance (squared magnitude of transmission coefficient) responses of the GRS-type and reference structures, as a function of the layer electrical thickness and the incidence angle, within the aforementioned parameter ranges. Specifically, Figs. \ref{Figure2}a--\ref{Figure2}c pertain to three representative GRS-type configurations at different stages of growth ($N=128, 512, 2048$ layers, respectively), whereas Figs. \ref{Figure2}d--\ref{Figure2}f and \ref{Figure2}g--\ref{Figure2}i show the corresponding results for the EMT predictions and periodic benchmarks, respectively. It is apparent the general agreement between the EMT and periodic cases, both displaying a rather marked transition from a highly transmissive (with mild Fabry-P\'erot-type oscillations) to an essentially opaque regime for incidence angles approaching the critical value ${\bar \theta}_c=60^o$. In fact, as previously mentioned, the EMT and periodic responses may differ substantially in the transition region around ${\bar \theta}_c$ \cite{Sheinfux:2014sm,Andryieuski:2015ae,Popov:2016oa,Lei:2017rt,Maurel:2018so,Castaldi:2018be,Gorlach:2019bc}, although this is hardly visible on the scale of the graphs here. Conversely, the responses of the GRS-type configurations are visibly different, with low-transmission regions extending for most of the parameter ranges, also far away from the critical-incidence condition. For increasing sizes, the high-transmission regions progressively reduce up to almost isolated points (see Fig. \ref{Figure2}c). This behavior is also markedly different from the fractal-bandgap structure observed for aperiodically ordered Thue-Morse-type geometries (compare with Figs. 2a-2c in \cite{Coppolaro:2018ao}).

%############################################################
%                Figure 4
%
\begin{figure}
	\centering\includegraphics[width=13.2cm]{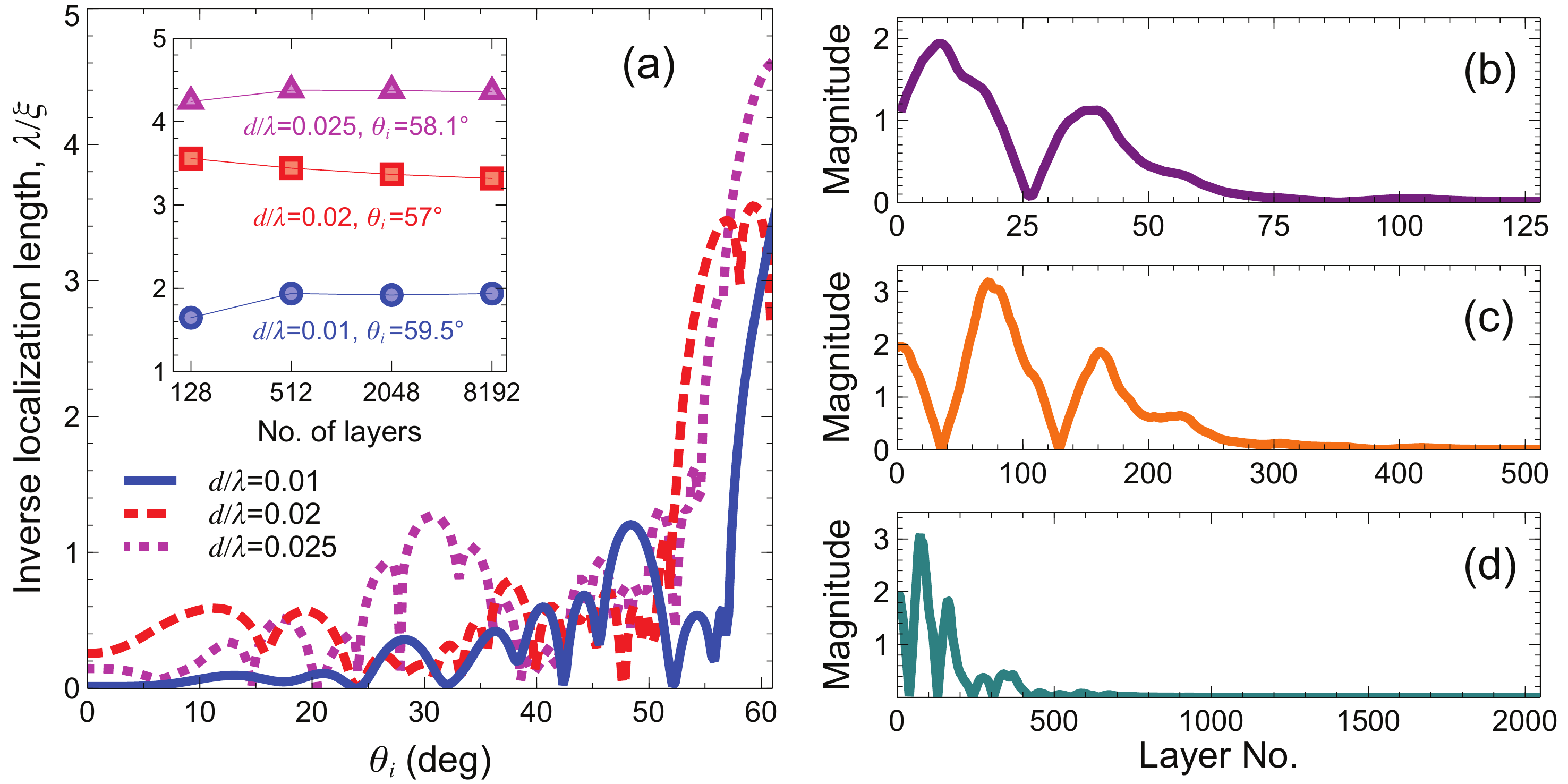}
	\caption{(a) Inverse localization length (scaled by the wavelength) for a GRS-type configuration with $N=2048$ layers and different electrical thicknesses (blue-solid: $d/\lambda=0.01$; red-dashed: $d/\lambda=0.02$; magenta-dotted: $d/\lambda=0.025$), as a function of the incidence angle. The inset shows the convergence to a constant value for increasing stages of growth and near-incidence conditions (continuous curves are guides to the eye only.).
		  (b), (c), (d) Representative electric-field (normalized-magnitude) distributions inside the GRS-type multilayer for  $N=128$ ($d/\lambda=0.02$, $\theta_i=58.35^o$),  $N=512$ ($d/\lambda=0.01$, $\theta_i=58.35^o$), and $N=2048$ ($d/\lambda=0.01$, $\theta_i=58.1^o$), respectively.}
	\label{Figure4}
\end{figure}
%############################################################

It is quite remarkable that, even for relatively small stages of growth (e.g., $N=128$ layers) and incidence conditions far from critical, the effect of deterministic disorder at deeply subwavelength scales can be so substantial. For instance, Fig. \ref{Figure3}a compares some representative cuts from Fig. \ref{Figure2} (for $N=128$ layers) for $\theta_i=45.1^o$, from which the differences between the widely oscillating GRS-type and the near-unity (hardly distinguishable) EMT and periodic cases are particularly evident. Figure \ref{Figure3}b shows the corresponding
internal electric-field (magnitude) distributions for $d/\lambda=0.01$, for which similar considerations hold.  We stress that the incidence angle is quite different from the critical value, and these structures are only wavelength-sized ($L=1.28\lambda$), with the smallest layer thickness considered in the study ($d/\lambda=0.01$). Even taking into account the maximum size of layer ``clusters'' (four consecutive identical symbols), the electrical size of these macro-regions is below one tenth of the minimum ambient wavelength, i.e., well within the deeply subwavelength regime.

The prominence of low-transmission regions in Figs. \ref{Figure2}a--\ref{Figure2}c indicates a tendency toward {\em localization} that somehow resembles what also observed for randomly disordered geometries \cite{Sheinfux:2016cr,Sheinfux:2017oo}. Although the localization-delocalization of states in systems characterized by deterministic disorder is still an unsettled issue \cite{Kroon:2002ld,Kroon:2004ao}, our observations seem to indicate that localization is indeed possible in the critical parameter regime of interest here.
For a more quantitative assessment, paralleling the studies on random geometries, it is expedient to define a {\em localization length} \cite{Sheinfux:2016cr,Sheinfux:2017oo}
\beq
\xi=-\frac{L}{\log T},
\eeq
where $T$ indicates the transmittance, and the ensemble average does not apply in view of the deterministic character of our geometry. For better visualization (in view of the more compressed dynamic range), Fig. \ref{Figure4}a shows the inverse localization length pertaining to a GRS-type configuration with $N=2048$ layers, as a function of the incidence angle (over an extended range $0<\theta_i\lesssim 60^o$), for three representative values of the layer electrical thickness. We observe a generally oscillatory behavior, with alternating peaks (i.e., minima of the localization length) and dips (indicative of high-transmission states), with an abrupt increase in the vicinity of the critical angle ${\bar \theta}_c=60^o$. Also shown, in the inset, is the behavior as a function of the stage of growth, for fixed electrical lengths and near-critical incidence conditions, from which we can observe the asymptotic convergence to a constant value. Overall, localization lengths of fractions of a wavelength are observed for near-critical incidence, but wavelength-sized values can also be observed far away from this region. Figures \ref{Figure4}b--\ref{Figure4}d show typical localized states for various stages of growth and near-critical incidence.

We highlight that the anomalies illustrated above are quite different from what observed in orderly (periodic or aperiodic) structures and, in spite of their fully deterministic nature, somehow more resemblant of the observations in randomly disordered geometries.

%############################################################
%                Figure 5
%
\begin{figure}
	\centering\includegraphics[width=7cm]{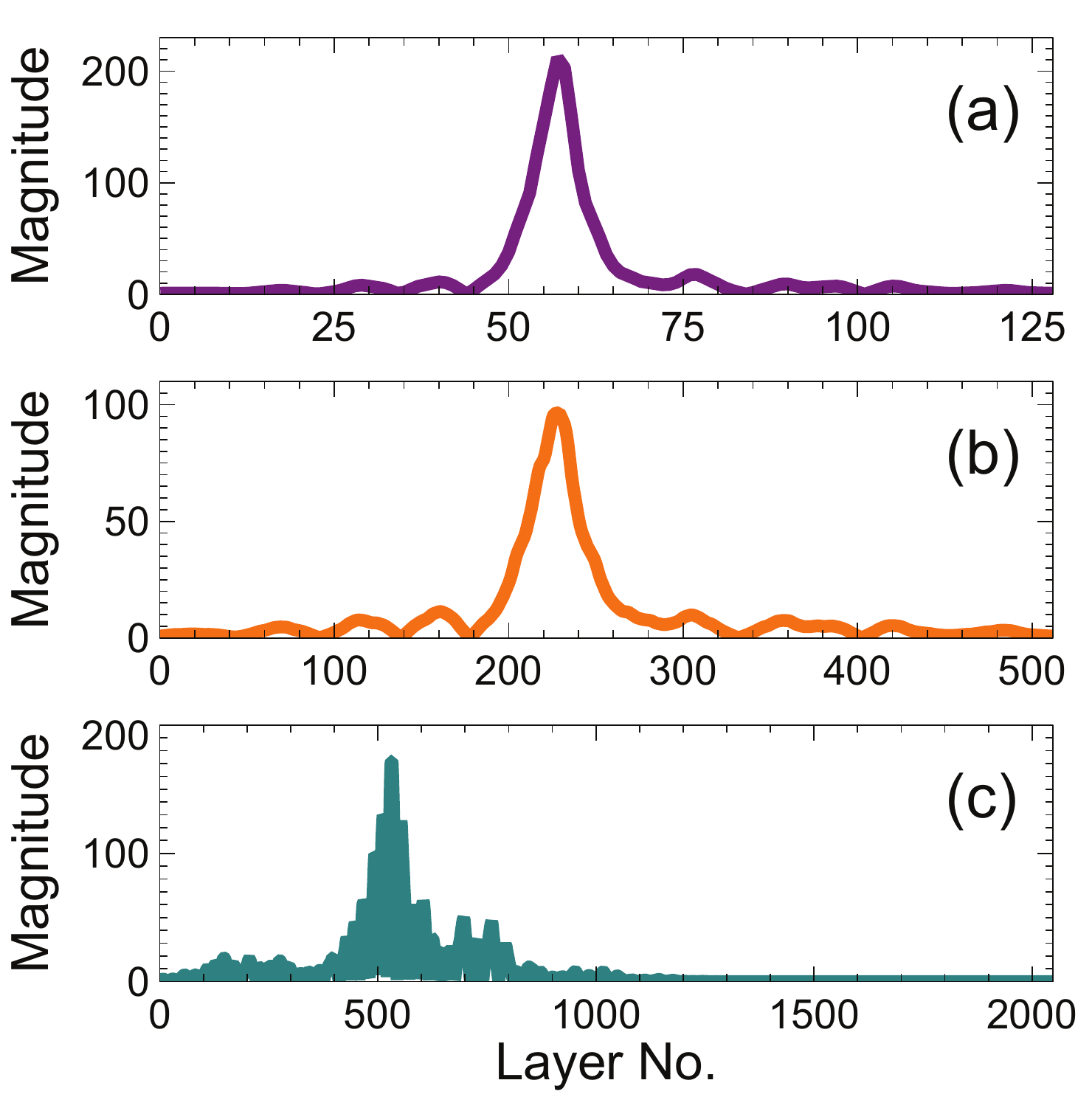}
	\caption{Representative electric-field (normalized-magnitude) distributions for localized states. (a) $N=128$, $d/\lambda=0.047$, $\theta_i=59.1^o$. (b) $N=512$, $d/\lambda=0.015$, $\theta_i=58.85^o$. (c) $N=2048$, $d/\lambda=0.02$, $\theta_i=46.6^o$.}
	\label{Figure5}
\end{figure}
%############################################################

%############################################################
%                Figure 6
%
\begin{figure}
	\centering\includegraphics[width=13.2cm]{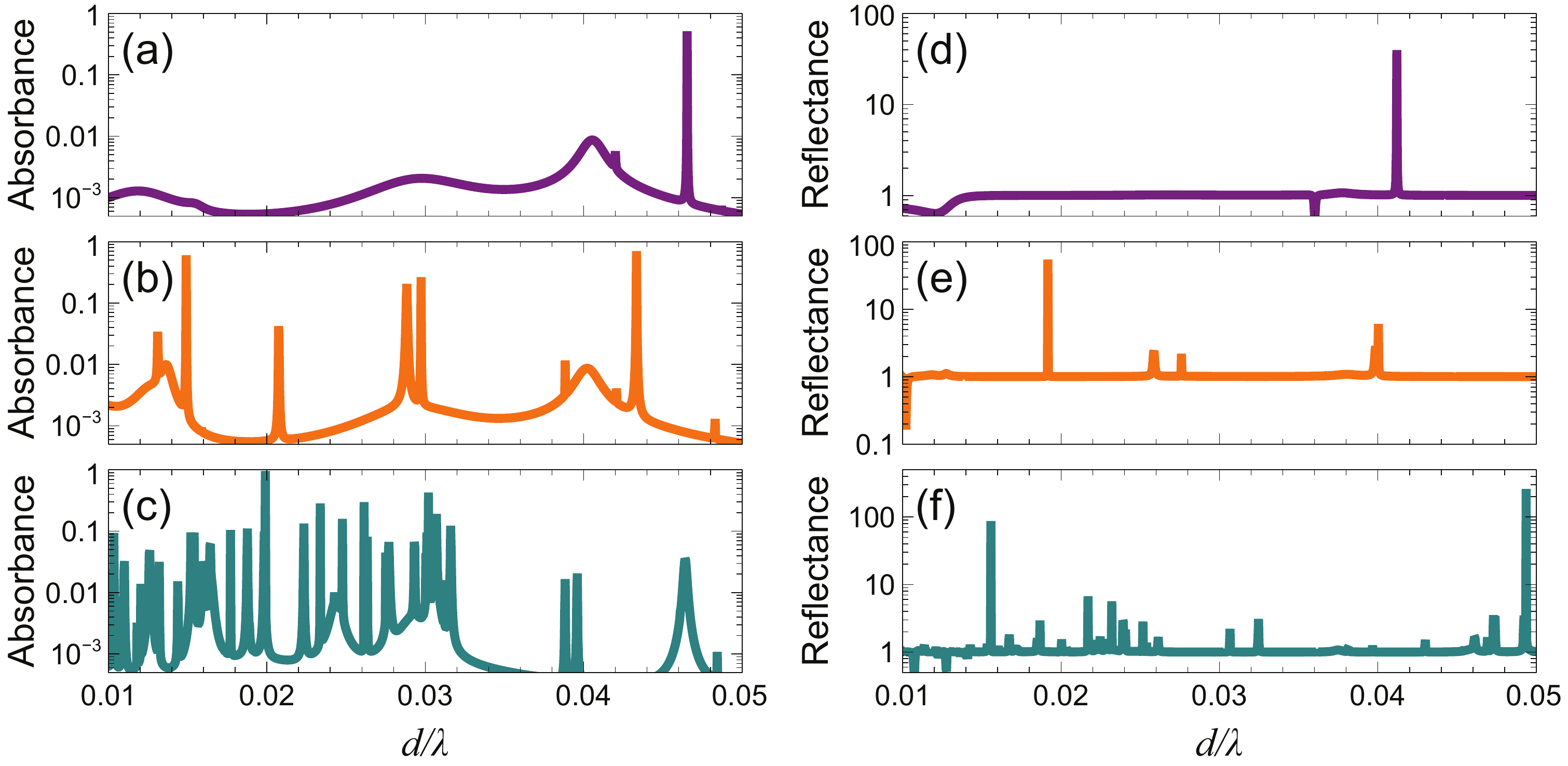}
	\caption{(a),(b),(c) Representative absorbance responses, in the presence of small losses ($\varepsilon_b=5+i10^{-4}$), as a function of electrical thickness, for $N=128$ ($\theta_i=59.1^o$), $N=512$ ($\theta_i=58.85^o$), and $N=2048$ ($\theta_i=46.6^o$), respectively. (d),(e),(f) Representative reflectance responses, in the presence of small gain ($\varepsilon_b=5-i10^{-3}$), as a function of electrical thickness, for $N=128$ ($\theta_i=56.85^o$), $N=512$ ($\theta_i=57.1^o$), and $N=2048$ ($\theta_i=36.1^o$), respectively.}
	\label{Figure6}
\end{figure}
%############################################################

%===============================================================================================
\subsection{Field enhancement, absorption and lasing}
%===============================================================================================
Another intriguing property of GRS-type multilayered metamaterials is their capability to support states featuring strong field enhancement, for specific parameter configurations. 
Figure \ref{Figure5} shows some representative results, selected from an extensive series of parametric studies, for different stages of growth and incidence conditions. We observe that enhancement factors of over two orders of magnitudes can be attained, also far away from the critical angle (see, e.g., Fig. \ref{Figure5}c). Similar enhancement phenomena have been observed in previous studies on GRS-based dielectric multilayers (see, e.g., \cite{Hiltunen:2007mo}), but in the photonic-crystal regime, i.e., for moderately sized layers. In fact, in the deeply subwavelength regime of interest here, the EMT predicts a field enhancement (see \cite{Coppolaro:2018ao} for details) 
$\bar{\gamma} ={\sqrt{\varepsilon_{e}} \cos \theta_{i}}/{\sqrt{\bar{\varepsilon}_{\|}-\varepsilon_{e} \sin ^{2} \theta_{i}}}$, which, for the parameters configurations in Fig. \ref{Figure6} ranges from $\sim 1.5$ to $\sim 4.4$ (i.e., up to two orders of magnitude smaller than the actual values observed). Results observed for the periodic benchmarks (not shown for brevity) are in line with the EMT predictions.

Intuitively, as also observed for the Thue-Morse aperiodically ordered case \cite{Coppolaro:2018ao}, such strong field enhancement can give rise to anomalous absorption or lasing in the presence of small losses or gain, respectively. Accordingly, we study the effects of a small imaginary part in the higher-permittivity material layers, $\varepsilon_b=5+i\delta$, where (in view of the assumed time-harmonic convention) positive and negative values of $\delta$ correspond to loss and gain, respectively. 

For different stages of growth and incidence conditions, Figs. \ref{Figure6}a--\ref{Figure6}c show the absorbance response as a function of the electrical thickness, by assuming very small losses ($\delta=10^{-4}$). As a reference, the EMT predictions and periodic benchmarks (not shown for brevity) are consistently below $\sim 10^{-2}$. As can be observed, GRS-type configurations can exhibit absorbances that are orders-of-magnitude higher than these predictions, with sharp peaks that can reach nearly unity, in  spite of the very small level of losses. The field distributions at these peaks (not shown for brevity) are qualitatively similar to those exhibiting field enhancement shown in Fig. \ref{Figure5}.

Finally, Figs. \ref{Figure6}d--\ref{Figure6}f show the reflectance responses in the presence of small gain ($\delta=-10^{-3}$). In this case, we observe the presence of sharp peaks with  strong amplitudes (up to $\sim100$), which indicate the onset of lasing conditions. These results confirm the possibility to significantly lower the lasing threshold, as an effect of deterministic disorder at deeply subwavelength scales. For instance, by comparing the results in Fig. \ref{Figure6}f (lasing peak at $d/\lambda=0.016$) with the lasing conditions attainable in the EMT scenario with same effective parameters (see \cite{Coppolaro:2018ao} for details), we estimate a reduction by a factor $\sim 50$ in the structure size (for fixed gain coefficient) or, equivalently, in the gain coefficient (for fixed structure size).

The above results may find potentially promising applications to absorbers and low-threshold lasers. 

%===============================================================================================
\subsection{Remarks}
%===============================================================================================
Results qualitatively similar to those illustrated above can also be observed for transverse-magnetic polarization and different material properties. However, as also found in the periodic case, their visibility is stronger for TE polarization and the stronger the material contrast \cite{Zhukovsky:2015ed}. Moreover, although our study was focused on $Q_{\nu}$-type (with odd $\nu$) GRS sequences, we have verified that qualitatively similar results (not shown for brevity) also hold for even values of $\nu$ as well as for $P_{\nu}$-type sequences (for either even or odd $\nu$), for which the statistical frequencies of occurrence of the symbols are exactly equal only in the asymptotic infinite-sequence limit.

Concerning the physical interpretation, similar to the periodic case \cite{Sheinfux:2014sm,Sheinfux:2016cr}, the buildup effects can be attributed to the peculiar (interface-dominated) phase-accumulation mechanism occurring in the structure. From the mathematical viewpoint, paralleling our previous studies on periodic and aperiodically ordered scenarios \cite{Castaldi:2018be,Coppolaro:2018ao},
the EMT-breakdown phenomenon can be parameterized as an {\em error-propagation} effect, whereby the elements of the exact and effective transfer matrix describing the light propagation in the structure progressively diverge for increasing stages of growth. In principle, also in this case such errors could be mitigated by introducing in the effective model suitable nonlocal corrections and magneto-electric coupling \cite{Castaldi:2018be,Popov:2016oa}. However, unlike the periodic case, here it is not possible to work out  some simple closed-form estimates for the error as well as the correction terms, since the mapping describing the transfer-matrix evolution is not solvable analytically. Similar to the periodic case, these anomalous effects become weaker for lower values of the material contrast, and tend to be especially emphasized for near-critical incidence conditions. However, as clearly visible in Fig. \ref{Figure2}, especially for higher stages of growth, the GRS geometry may exhibit much stronger departures from the EMT predictions also considerably far from critical incidence. Within this framework, a systematic study of the trace and anti-trace maps \cite{Wang:2000ta} of the transfer matrix may provide some useful insights.

As for a possible experimental verification, although this is not the focus of the present study, we point out that the constitutive, geometrical and excitation conditions are not far from realistic. In particular, although our study also includes results for unrealistically large (thousands-layer) structures, certain effects are significantly visible  also for realistic sizes of hundred(s) layers.
We remark that our choice of the permittivity values was essentially motivated by facilitating direct comparisons with previous theoretical studies on periodic, aperiodically ordered, and random scenarios in Refs. \cite{Sheinfux:2014sm,Sheinfux:2016cr,Coppolaro:2018ao}, where the same values were considered. It is also worth emphasizing that, as experimentally demonstrated in Ref. \cite{Sheinfux:2017oo} for random scenarios, the anomalous localization phenomena remain visible in the presence of realistic materials (silica and niobium pentoxide for the layers, and rutile for the illumination prism).

%%%%%%%%%%%%%%%%%%%%%%%%%%%%%%%%%%%%%%%%%%%%%%%%%%%%%%%%%%%%%%
\section{Conclusions}
%%%%%%%%%%%%%%%%%%%%%%%%%%%%%%%%%%%%%%%%%%%%%%%%%%%%%%%%%%%%%%
\label{Sec:Conclusions}
In conclusion, we have shown that finite-size multilayered dielectric metamaterials can exhibit enhanced sensitivity to deterministic disorder at deeply subwavelength scales.
With specific reference to a simple GRS-type model, we have studied in detail these effects, by identifying certain critical parameter regimes and manifestations in terms of significant departures of the optical response (in terms of transmittance, localization, field-enhancement and absorption/lasing) from the EMT predictions, and have illustrated the underlying mechanisms. 

Our results enrich and complement previous studies on periodic and aperiodic (orderly and random) scenarios, indicating some common traits but also key differences that are distinctive of deterministic disorder. Besides the inherent implications for homogenization theory, the enhanced sensitivity to spatial (dis)order at deeply subwavelength scales 
may open new pathways for optical sensing, switching and lasing applications. Current and future studies are aimed at exploring these applications as well as different geometrical arrangements, also in higher-dimensional configurations. Of particular interest would be geometrical arrangements that could be parameterized so as to study the transition from orderly to disordered scenarios. Furthermore, a comparison between the effects induced by deterministic and random disorder could also provide very useful insights.
Finally, along the lines of recent studies on random geometries \cite{Dikopoltsev:2019co,Sharabi:2019lp}, exploring the effects of deterministic (dis)order in non-Hermitian and  time-varying multilayers also appears of great interest.

\section*{Funding}
University of Sannio (FRA program).

\section*{Disclosures}
The authors declare no conflicts of interest.

%%%%%%%%%% If using BibTeX:
%\bibliography{RS-ML.bib}

%%%%%%%%%% If preparing manually:
% \newcommand{\enquote}[1]{``#1''}

% \bibitem{Zhang:14}
% Y.~Zhang, S.~Qiao, L.~Sun, Q.~W. Shi, W.~Huang, L.~Li, and Z.~Yang,
%   \enquote{Photoinduced active terahertz metamaterials with nanostructured
%   vanadium dioxide film deposited by sol-gel method,}
%   {\protect\JournalTitle{Optics Express}} \textbf{22}, 11070--11078 (2014).

% {Optical Society}, \enquote{{OSA Publishing},}
%   \url{http://www.osapublishing.org}.

% \bibitem{FORSTER2007}
% P.~Forster, V.~Ramaswamy, P.~Artaxo, T.~Bernsten, R.~Betts, D.~Fahey,
%   J.~Haywood, J.~Lean, D.~Lowe, G.~Myhre, J.~Nganga, R.~Prinn, G.~Raga,
%   M.~Schulz, and R.~V. Dorland, \enquote{Changes in atmospheric consituents and
%   in radiative forcing,} in \enquote{Climate Change 2007: The Physical Science
%   Basis. Contribution of Working Group 1 to the Fourth assessment report of
%   Intergovernmental Panel on Climate Change,}  S.~Solomon, D.~Qin, M.~Manning,
%   Z.~Chen, M.~Marquis, K.~B. Averyt, M.~Tignor, and H.~L. Miler, eds.
%   (Cambridge University Press, 2007).

% \end{thebibliography}

\end{document}